\documentclass{PoS}

\title{SUSY Flavour at LHC7}

\ShortTitle{SUSY Flavour at LHC7}

\author{\speaker{J.~Jones-P\'erez}\\
        INFN, Laboratori Nazionali di Frascati, Via E.~Fermi 40, I-00044 Frascati, Italy\\
        E-mail: \email{joel.jones@lnf.infn.it}}

\author{L.~Calibbi\\
        Max-Planck-Institut f\"ur Physik (Werner-Heisenberg-Institut), F\"ohringer Ring 6, D-80805 M\"unchen, Germany\\
        E-mail: \email{calibbi@mppmu.mpg.de}}

\author{R.~N.~Hodgkinson\\
        Departament de F\'{\i}sica Te\`orica and IFIC, Universitat de Val\`encia-CSIC, E-46100, Burjassot, Spain\\
        E-mail: \email{robert.hodgkinson@uv.es}}

\author{A.~Masiero\\
        Dipartimento di Fisica, Universit\`a di Padova, via F.~Marzolo 8, I--35131, Padova, Italy\\
	INFN, Sezione di Padova, via F.~Marzolo 8, I--35131, Padova, Italy\\
        E-mail: \email{antonio.masiero@pd.infn.it}}

\author{O.~Vives\\
        Departament de F\'{\i}sica Te\`orica and IFIC, Universitat de Val\`encia-CSIC, E-46100, Burjassot, Spain\\
        E-mail: \email{oscar.vives@uv.es}}

\abstract{The current 7 TeV run of the LHC experiment shall be able to probe gluino and squark masses up to values of about 1 TeV. Assuming that hints for SUSY are found by the end of a 2 fb$^{-1}$ run, we explore the flavour constraints on the parameter space of the CMSSM, with and without massive neutrinos. In particular, we focus on decays that might have been measured by the time the run is concluded, such as $B_s\to\mu\mu$ and $\mu\to e\gamma$. We also briefly show the impact such a collider--flavour interplay would have on a Flavoured CMSSM.}

\FullConference{The 2011 Europhysics Conference on High Energy Physics, EPS-HEP 2011,\\
		July 21-27, 2011\\
		Grenoble, Rh\^one-Alpes, France}

\begin{document}

\section{Introduction}

With the LHC currently testing SUSY models, one can ask if flavour data can provide any additional information if any hints for SUSY are found. This work presents an attempt to find such a collider-flavour interplay, based on the data available before July, 2011. For an updated version of this work, with a luminosity of up to 5 fb$^{-1}$, the reader should refer to~\cite{Calibbi:2011dn}.

In the following, we shall study flavour observables within three models. The first model of interest is the CMSSM, which provides a standard benchmark. As usual, we shall scan the standard parameters: $m_0$, $M_{1/2}$, $A_0$ and $\tan\beta$, with ${\rm sgn}(\mu)>0$. Here, the contributions to flavour observables follow a MFV pattern, generated through RGE running. The second model of interest is a CMSSM with additional right-handed neutrinos. We assume scenarios with CKM- and PMNS-like mixings in the $Y_\nu$ matrices, generating LFV processes through the SUSY-Seesaw~\cite{Masiero:2002jn}. The final model of interest is a Flavoured CMSSM, in particular, the second model of~\cite{Calibbi:2009ja}. Here, flavour structures for the soft-breaking terms are generated based on an SU(3) flavour symmetry that simultaneously generates the observed flavour structures in the Yukawa matrices of the SM.

\section{LHC + Flavour Interplay}

\subsection{The LHC Reach with 7 TeV and 2 fb$^{-1}$}

At the time this work was presented, the most stringent constraints on the CMSSM parameter space were based on the ATLAS analysis of multi-jet events with missing energy and no leptons in the final state, using 35 pb$^{-1}$~\cite{daCosta:2011qk}. This constrained the $m_0$--$M_{1/2}$ plane with practically no restrictions on neither $\tan\beta$ nor $A_0$. Also, the analysis of~\cite{Baer:2010tk} determined the future reach of the experiment with a 2 fb$^{-1}$ luminosity, which was roughly the expected luminosity for the end of 2011. Thus, both analysis determined a band on the $m_0$--$M_{1/2}$ plane for this run where SUSY could be probed.

In order to carry out our analysis, we performed a scan within this band, scanning freely the other parameters: $\tan\beta\in[5,60]$ and $a_0=A_0/m_0\in[-3,3]$. This was carried out with SPheno~\cite{Porod:2011nf}.

\subsection{CMSSM}

\begin{figure}[p]
\begin{center}
\includegraphics[width=0.3\textwidth]{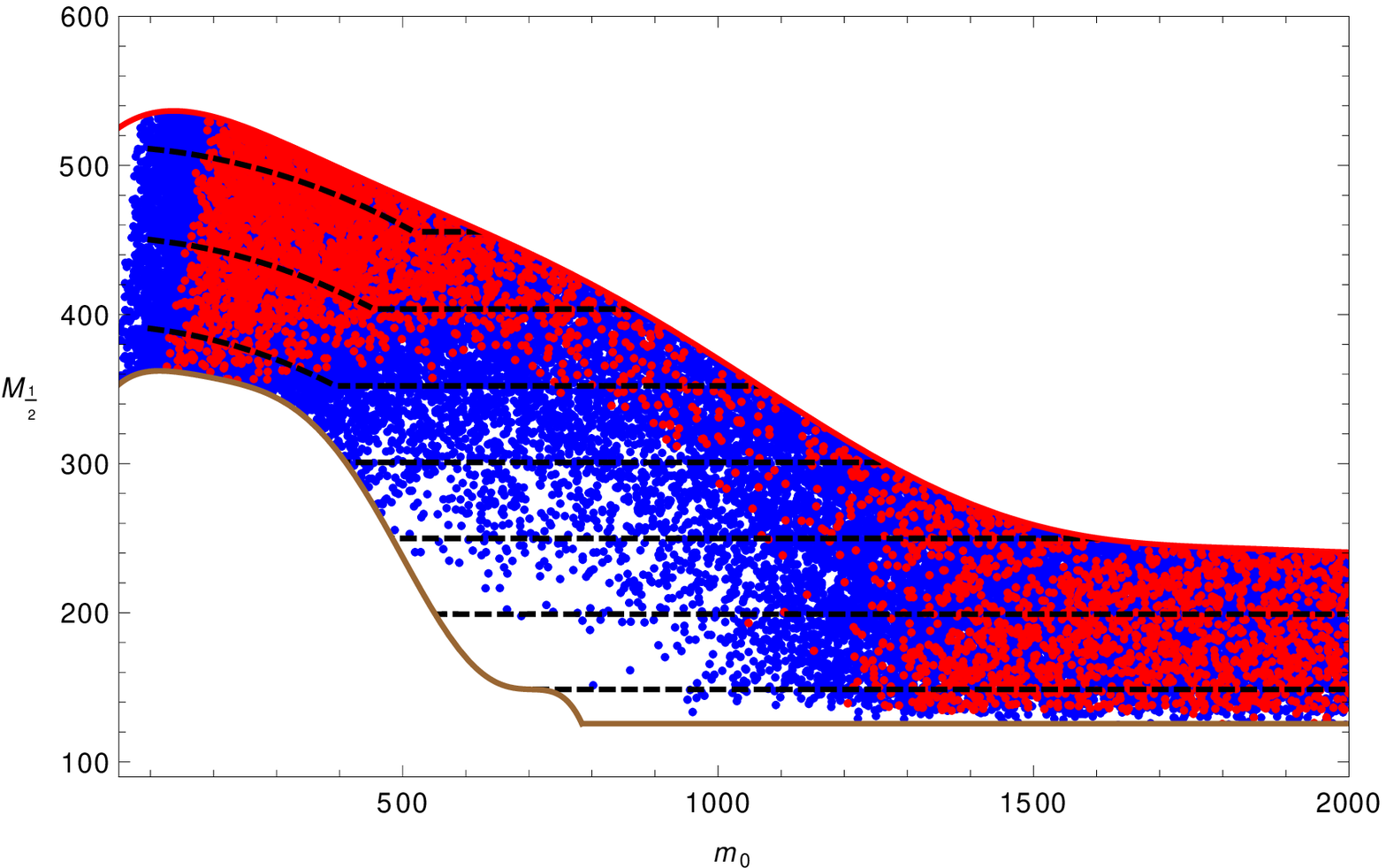}
\includegraphics[width=0.3\textwidth]{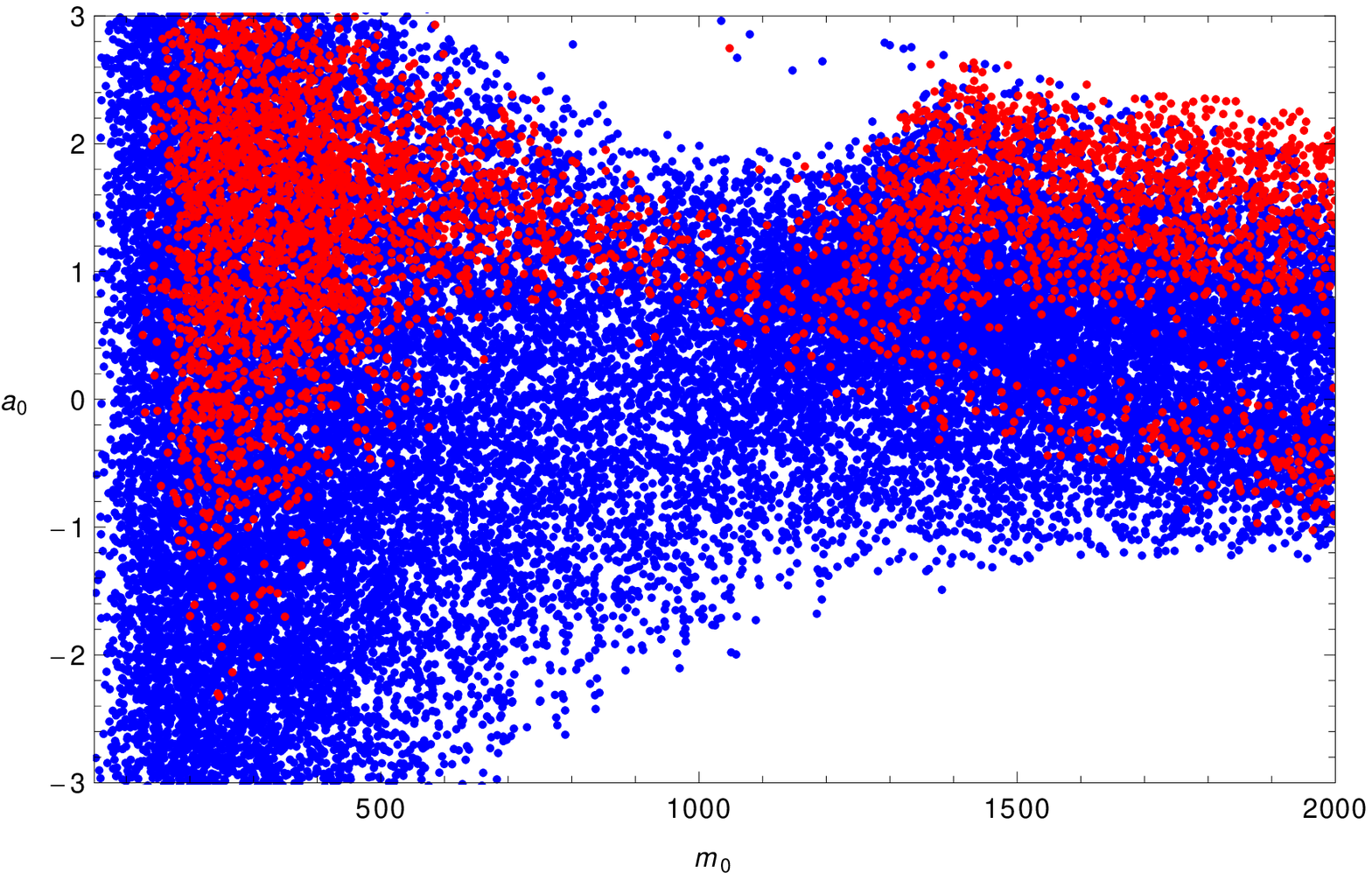}
\includegraphics[width=0.3\textwidth]{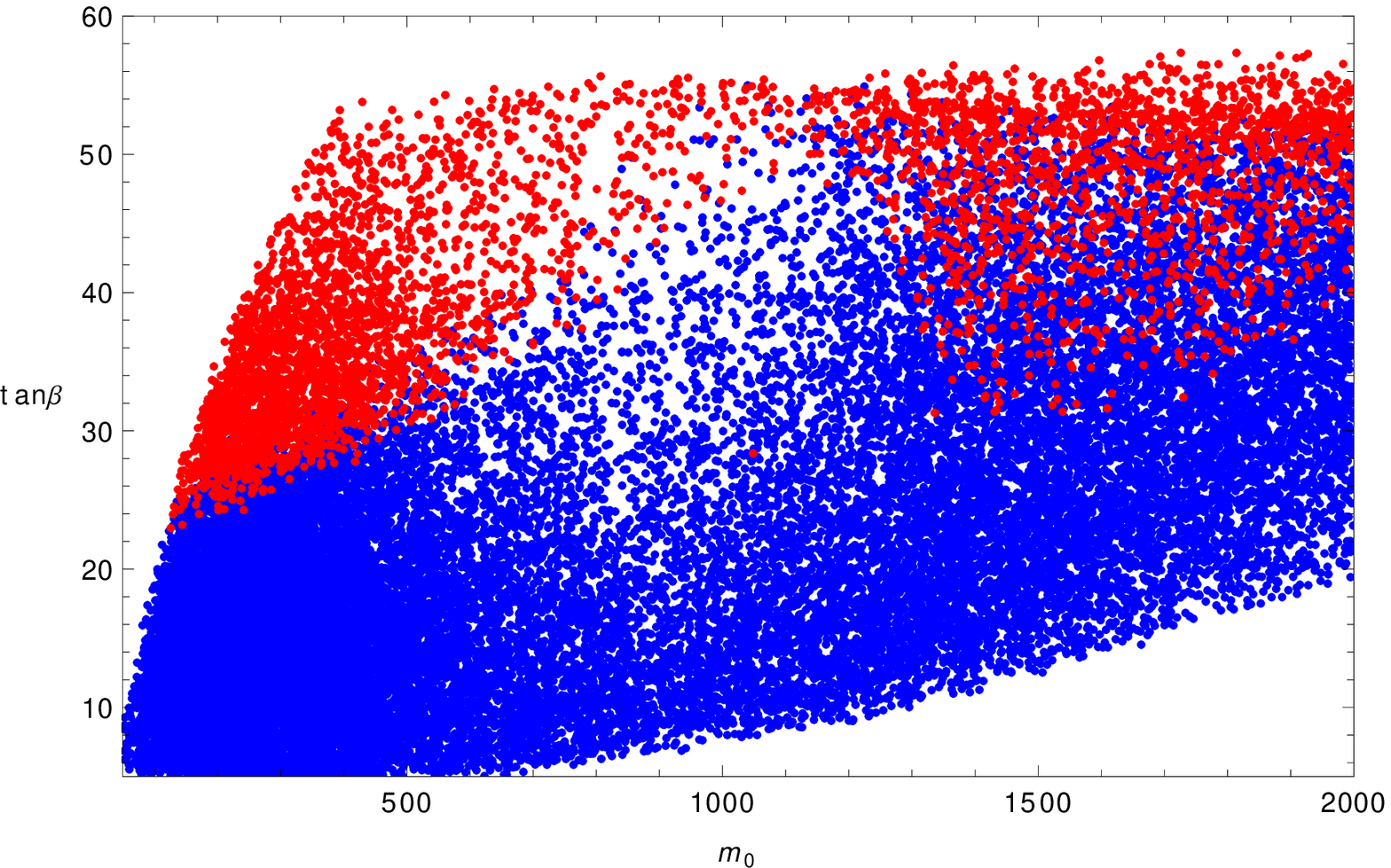} \\
\includegraphics[width=0.3\textwidth]{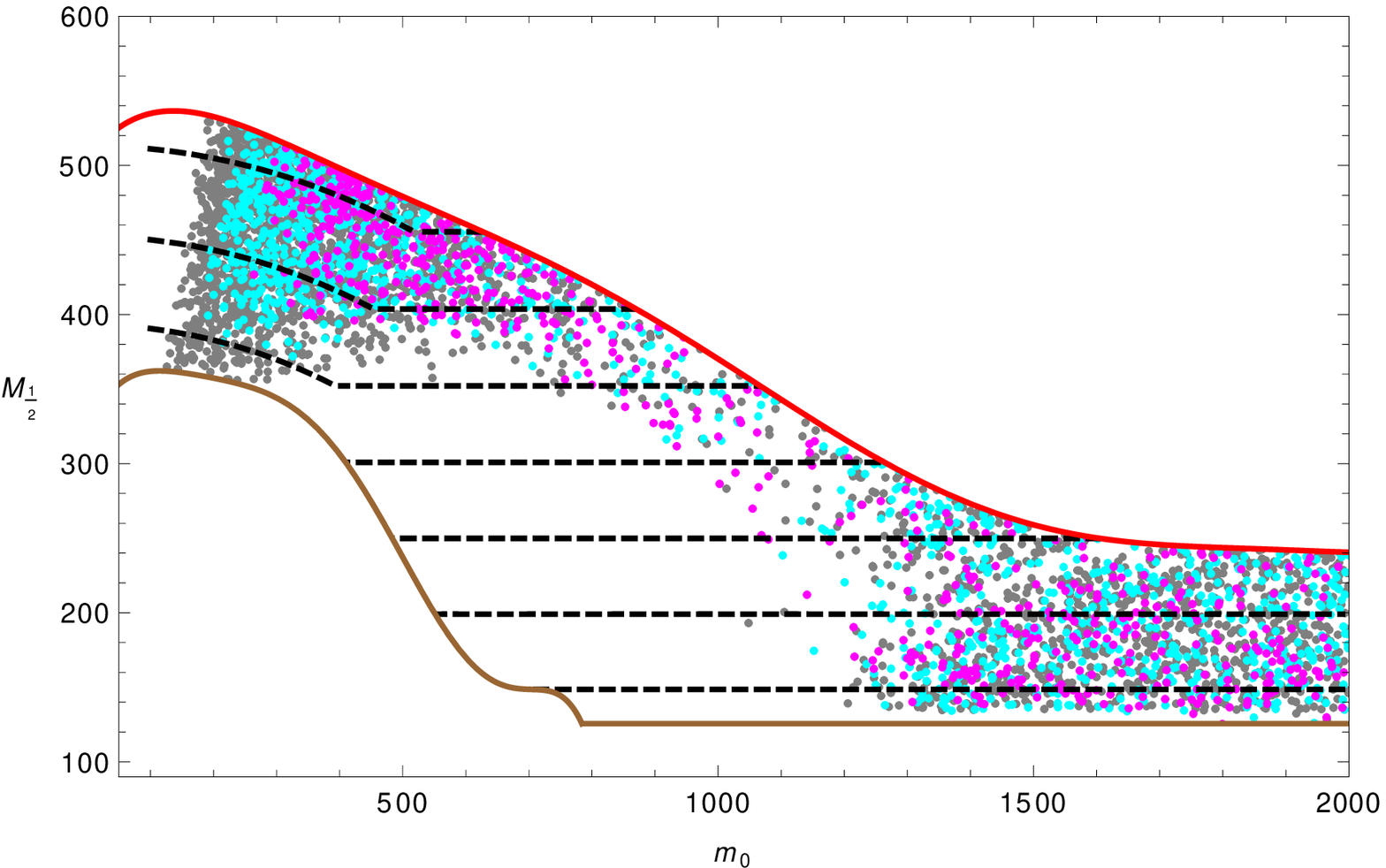}
\includegraphics[width=0.3\textwidth]{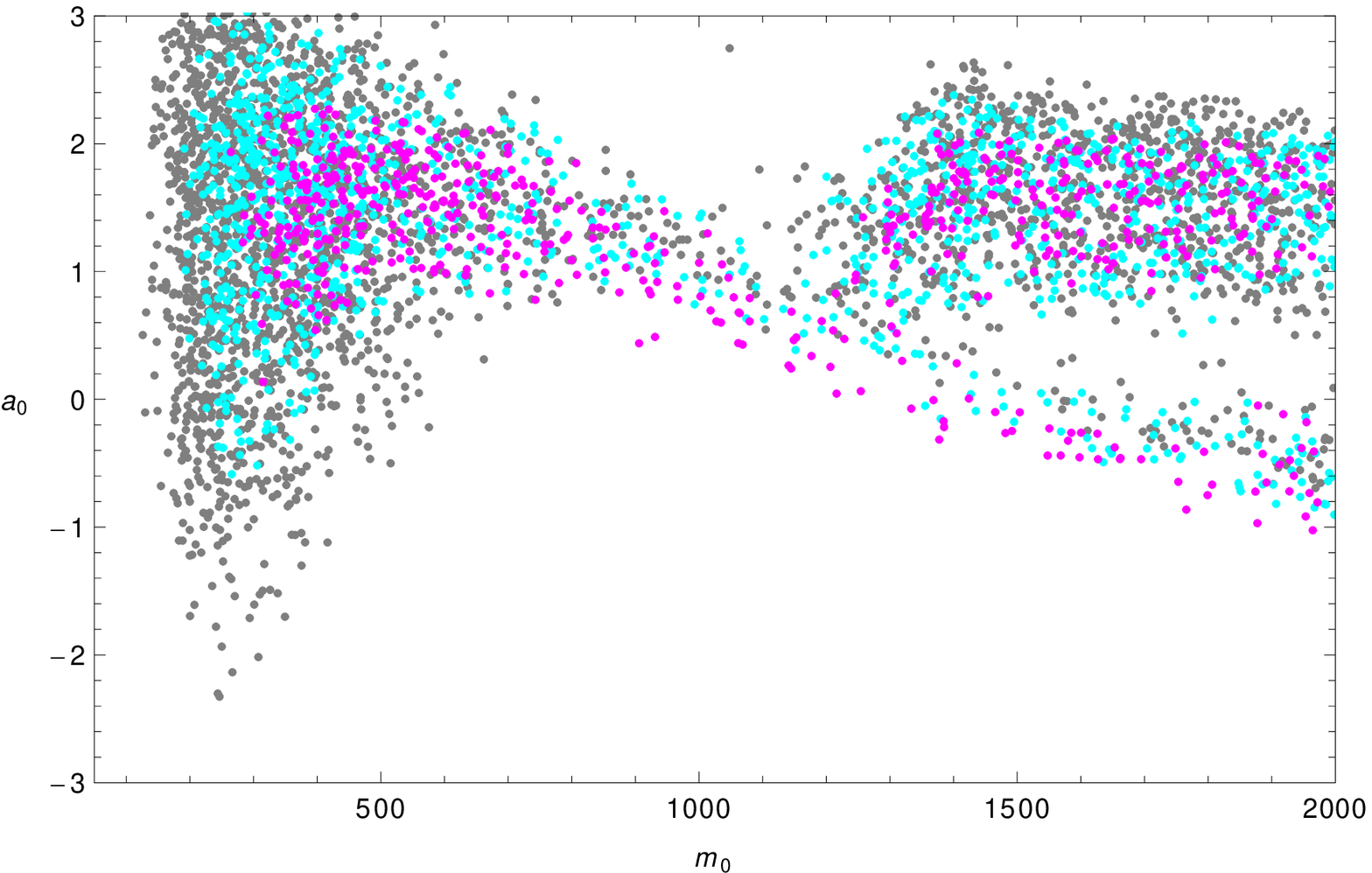}
\includegraphics[width=0.3\textwidth]{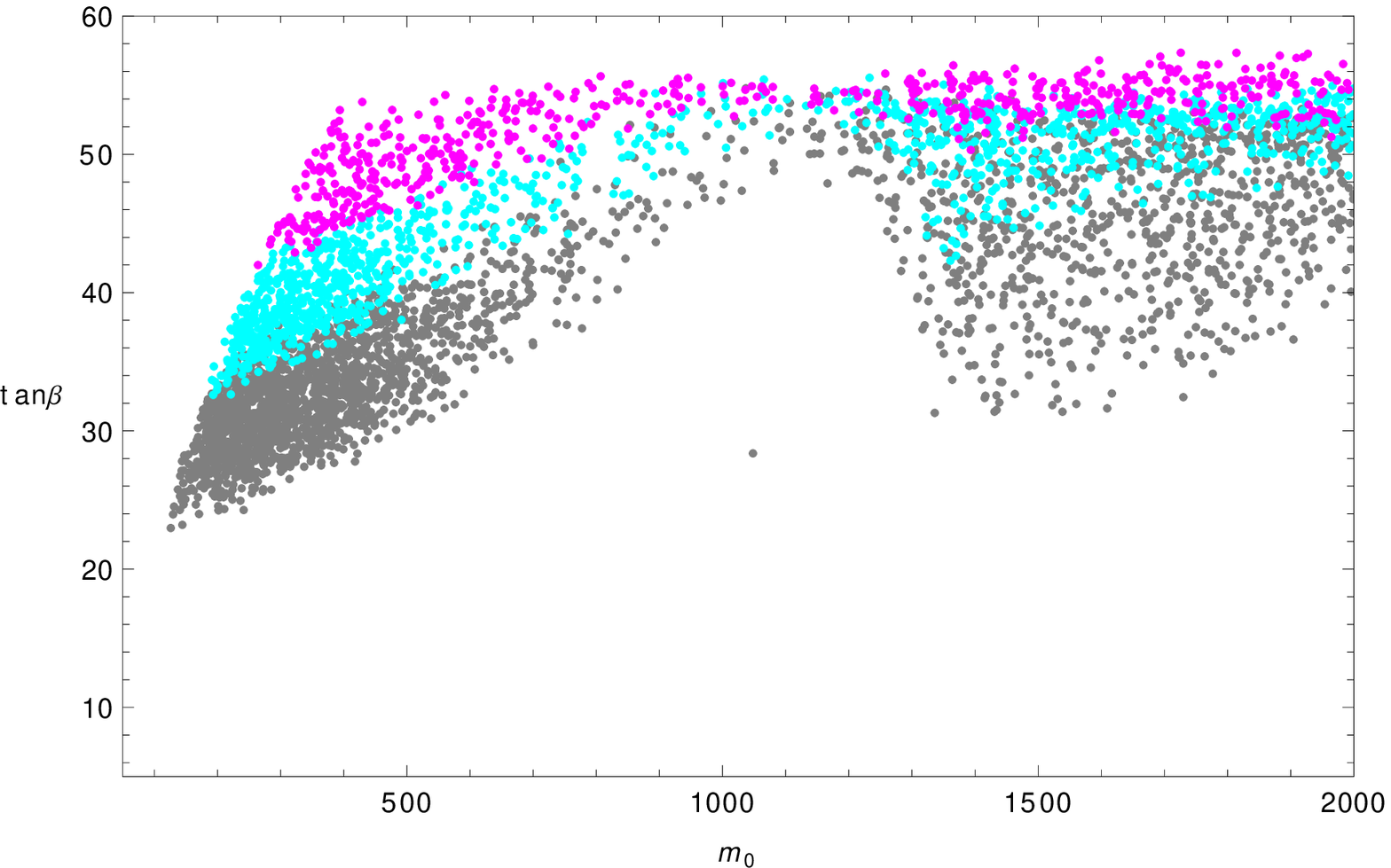}
\end{center}
\caption{Upper row: CMSSM parameter space allowed by $3\sigma$ flavour bounds and direct searches. Red points have $BR(B_s\to\mu^+\mu^-)>4\times10^{-9}$. Lower row: Points with a large $BR(B_s\to\mu^+\mu^-)$ giving a $5\sigma$ discovery ($3\sigma$ evidence) at LHCb are shown in magenta (cyan). Points in grey would give a signal under $3\sigma$.}
\label{fig:bsmumu1}
\end{figure}

We apply constraints on the scan based on achieving EW symmetry breaking, and the non-existance of a charged LSP, tachyons nor tensions with flavour data of more than $3\sigma$. Of the latter, the strongest constraints were due to $b\to s\gamma$ decay and $(g-2)_\mu$. The points in Figure~\ref{fig:bsmumu1} satisfy these constraints, where we find regions in the evaluated planes that are completely ruled out.

An observable of interest at this time is $BR(B_s\to\mu^+\mu^-)$. The LHCb collaboration claims that, with 2~fb$^{-1}$ of data, it can achieve a $5\sigma$ discovery of $BR(B_s\to\mu^+\mu^-)\gtrsim9\times10^{-9}$, or find $3\sigma$ evidence of $BR(B_s\to\mu^+\mu^-)\gtrsim5\times10^{-9}$. In the case of not seeing any signal, the same experiment claims to be able to rule out any branching ratio larger than $4\times10^{-9}$ with 95\% confidence~\cite{LambertMoriond}. Thus, we can ask what would be the consequences if this decay is or is not observed.

The red points in the upper part of Figure~\ref{fig:bsmumu1} show cases with $BR(B_s\to\mu^+\mu^-)>4\times10^{-9}$, i.e.\ those to be discarded if LHCb does not see anything. We find that $B_s\to\mu^+\mu^-$ can put strong bounds on the parameter space. In addition, in the lower part of the Figure, we show in cyan (magenta) those points that would give a $5\sigma$ ($3\sigma$) signal. Such an observation would also constrain significantly the parameter space, especially in the $\tan\beta$--$m_0$ and $a_0$--$m_0$ planes.

\subsection{CMSSM + $\nu_R$}

\begin{figure}[p]
\begin{center}
\includegraphics[angle=-90, width=0.3\textwidth]{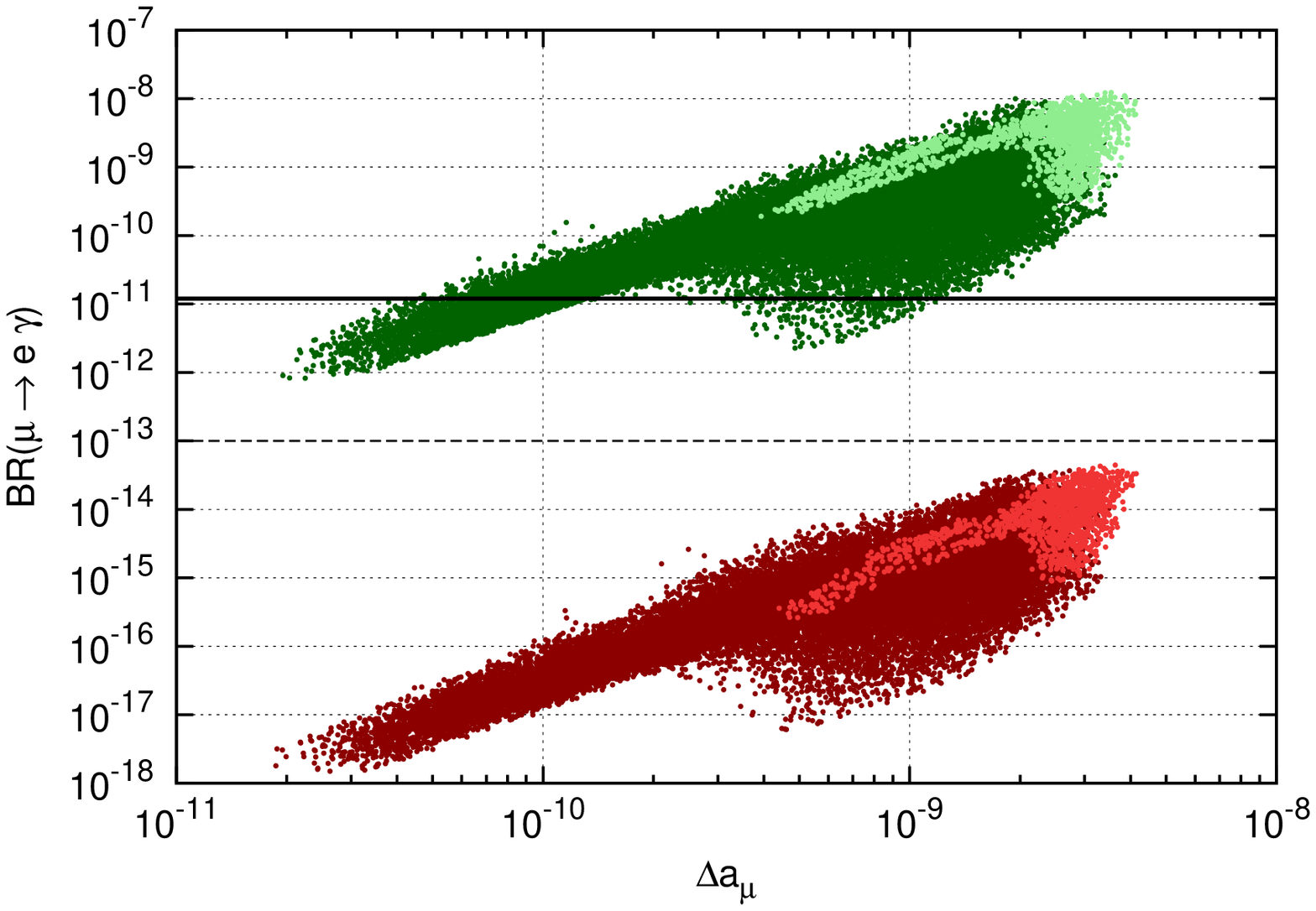}
\includegraphics[angle=-90, width=0.3\textwidth]{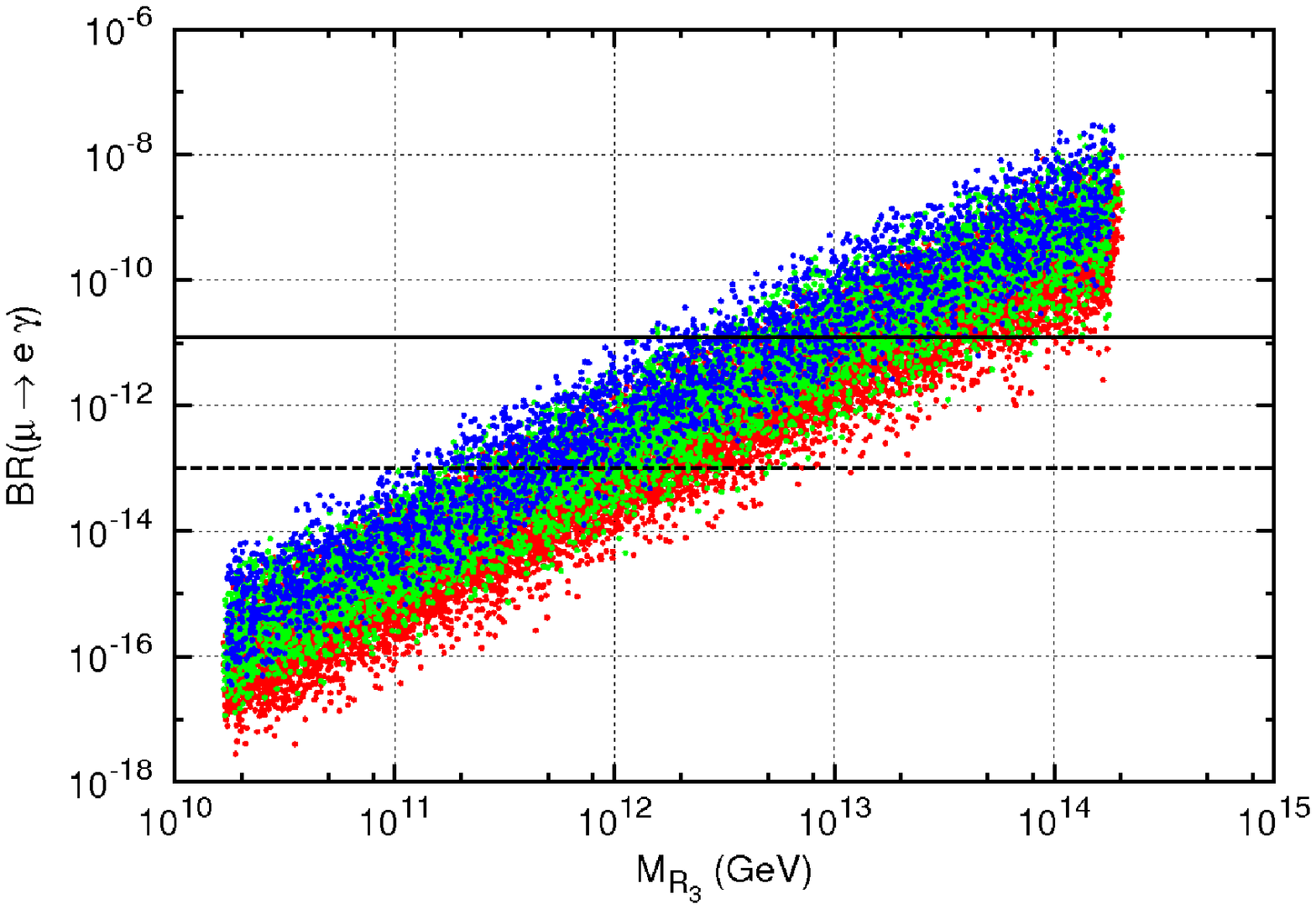}
\includegraphics[angle=-90, width=0.3\textwidth]{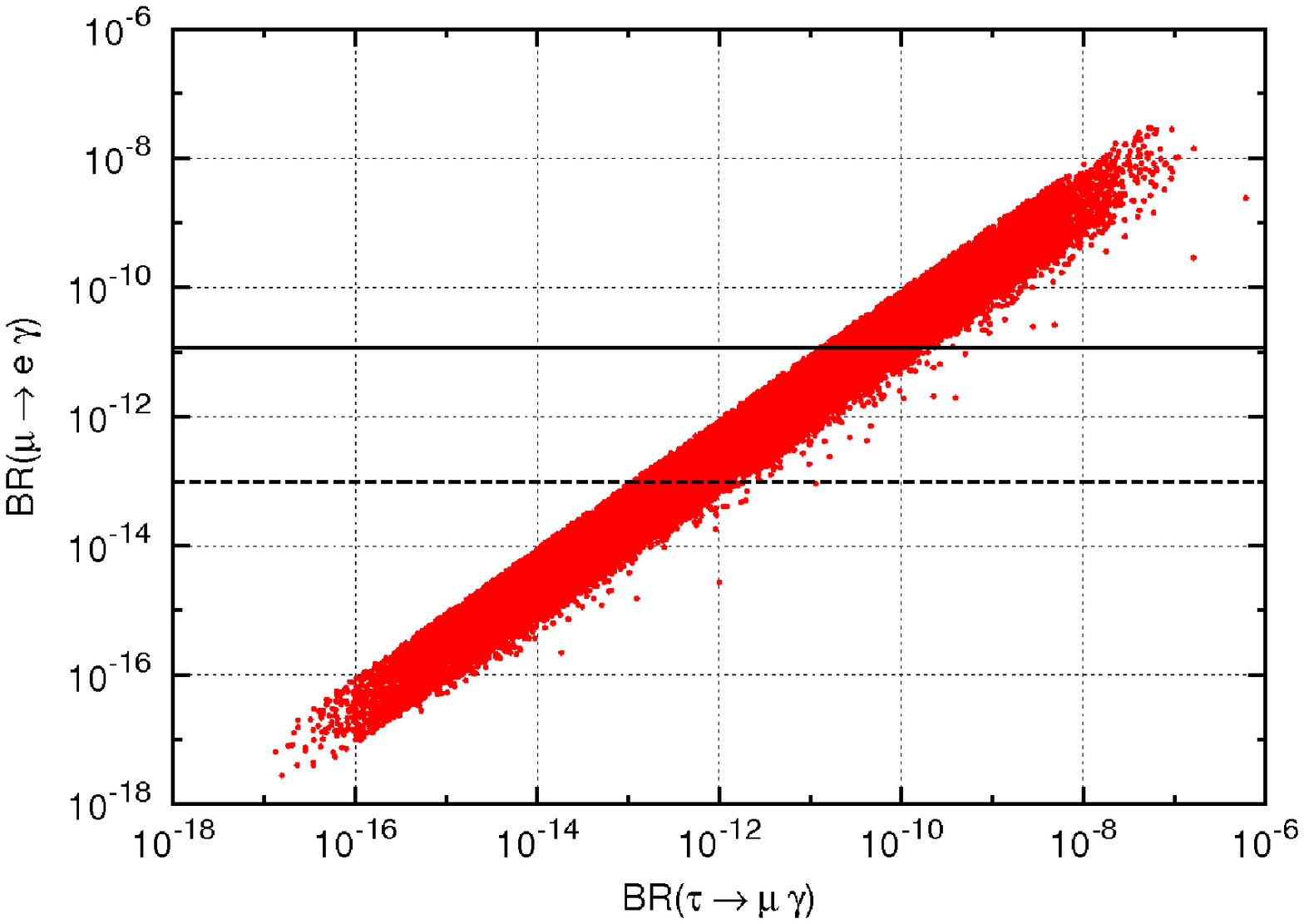}
\end{center}
\caption{CMSSM + $\nu_R$ phenomenology. $BR(\mu\to e\gamma)$ vs $\Delta a_\mu$ (left), $M_{R_3}$ (center) and $BR(\tau\to\mu\gamma)$ (right). On the left, green (red) points correspond to PMNS (CKM) mixing, with the lighter points giving at least a $3\sigma$ $BR(B_s\to\mu^+\mu^-)$ signal. On the center, green (blue) points give $\delta a_u>10^{-9}$ ($2\times10^{-9}$). In all plots, solid (dashed) lines show current (future) bounds.}
\label{fig:NuR}
\end{figure}

The main observables in a SUSY SeeSaw are LFV processes. In the left panel of Figure~\ref{fig:NuR} we show the state of $BR(\mu\to e\gamma)$ in the CKM- and PMNS-mixing scenarios. We find that the PMNS-mixing is practically ruled out by the MEGA constraints if we demand the $(g-2)_\mu$ tension to be solved. In contrast, the CKM-mixing cannot be probed, not even by MEG.

In the central panel of Figure~\ref{fig:NuR} we show a special case of PMNS-mixing, with the possibility of small Yukawa couplings. Here, the $BR(\mu\to e \gamma)$ bound constrains the heaviest RH neutrino mass. We find that, in this situation, the MEG experiment can probe masses down to $10^{13}$ GeV.

Finally, on the right panel, we show the correlation between $\mu\to e\gamma$ and $\tau\to\mu\gamma$ in the latter scenario. We find that, if the $\mu\to e\gamma$ bounds are to be fulfilled, then it shall not be feasible to observe a $\tau\to\mu\gamma$ signal larger than $10^{-9}$, which is the best expectation for Super Flavour Factories.

\subsection{Flavoured CMSSM}

The considered model includes additional flavour and CPV structures. Thus, they are more strongly bounded by FCNC than MFV-like cases, but also have a more interesting phenomenology.

\begin{figure}[p]
\begin{center}
\includegraphics[width=0.3\textwidth]{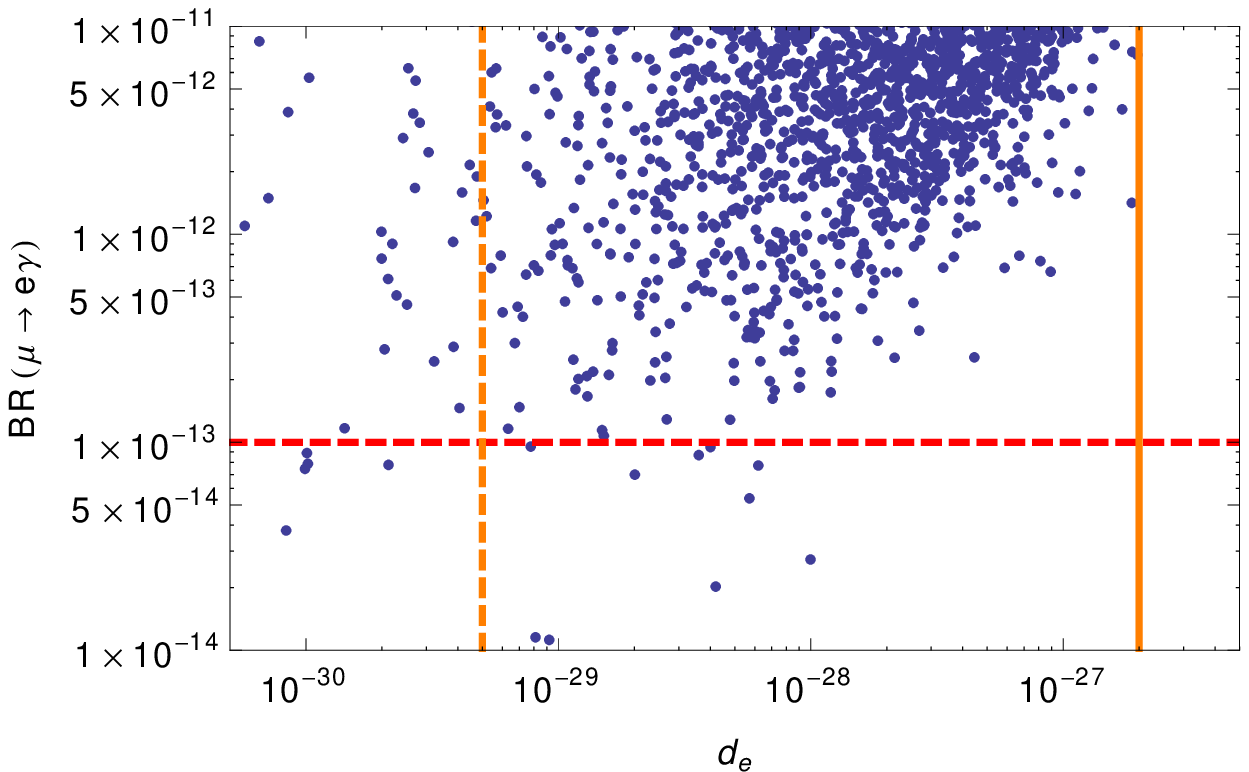}
\includegraphics[width=0.3\textwidth]{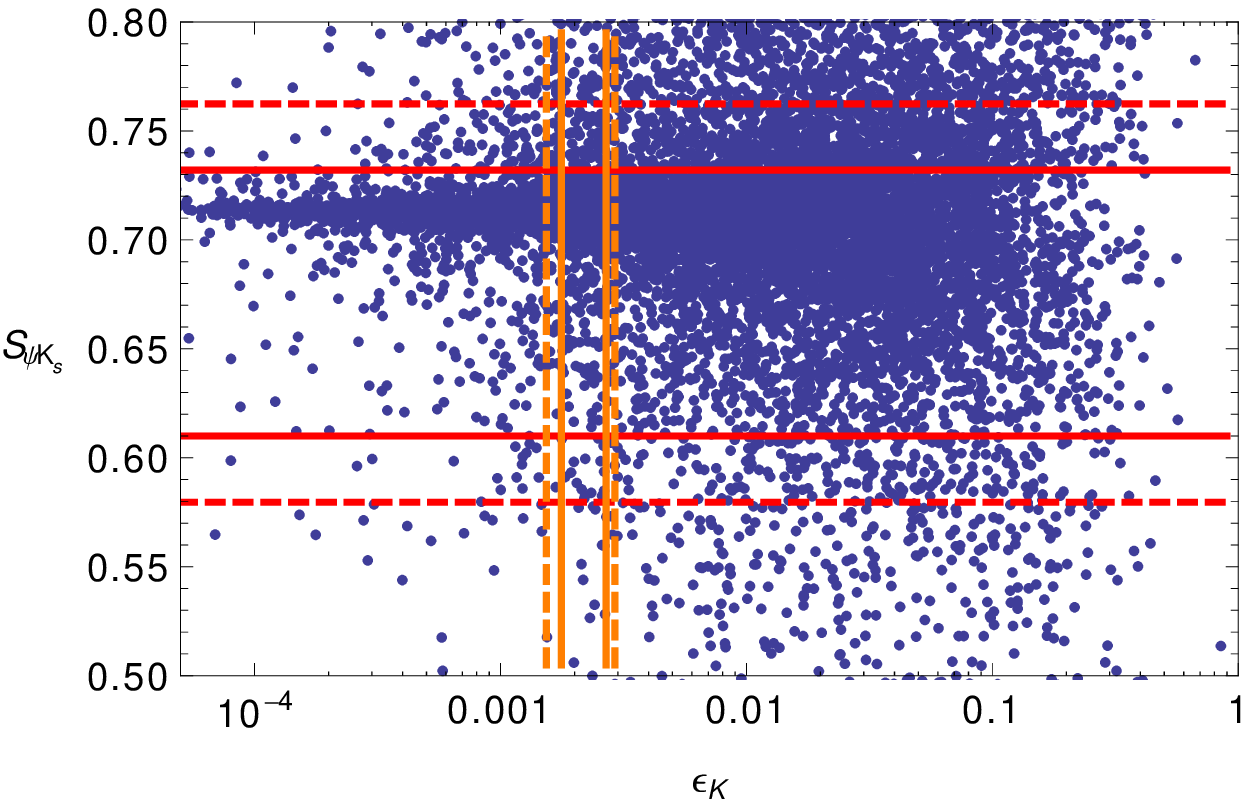}
\includegraphics[width=0.3\textwidth]{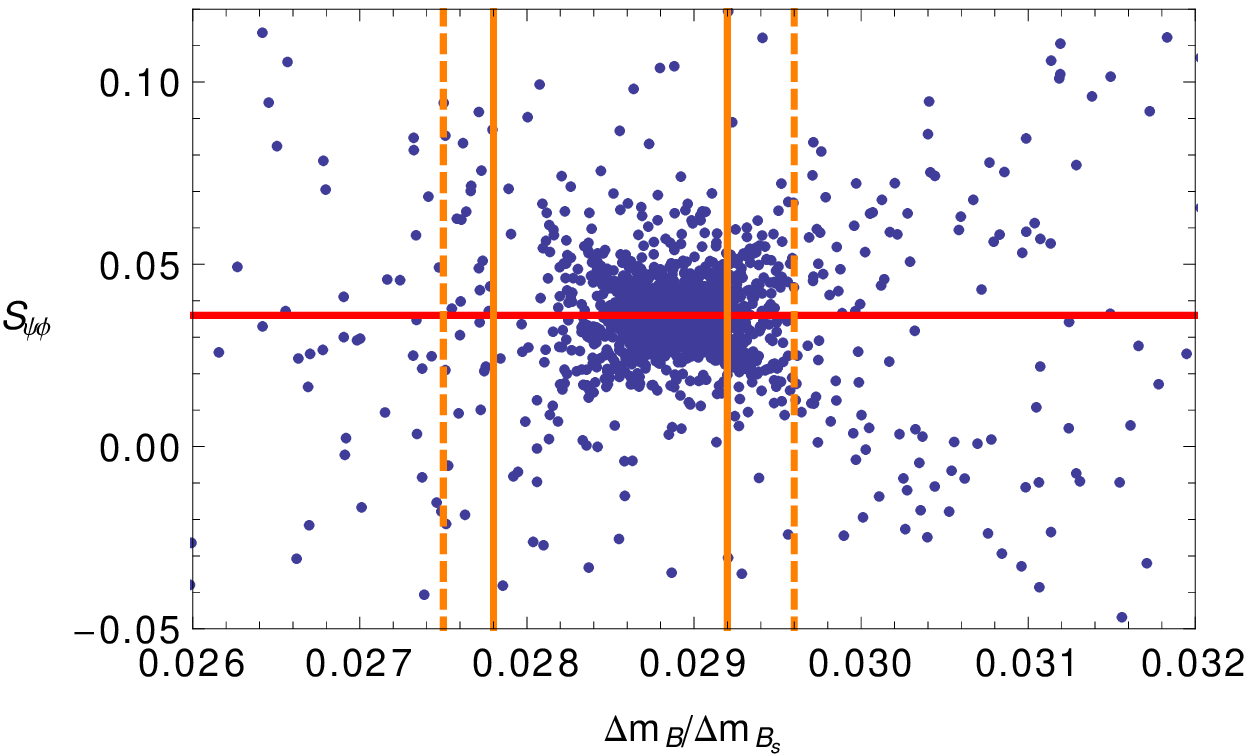}
\end{center}
\caption{RVV2 phenomenology. Left: $BR(\mu\to e\gamma)$ vs $d_e$. Red line gives MEG prospect for 2011 data, solid (dashed) orange line gives $d_e$ current bound (future prospects). Center: Constraints by $S_{\psi K_s}$ and $\epsilon_K$. Solid (dashed) lines give $2\sigma$ ($3\sigma$) bounds. Right: $S_{\psi\phi}$ prediction vs $\Delta m_B/\Delta m_{B_s}$ bound. Red line give central value of SM prediction, solid (dashed) lines give experimental $2\sigma$ ($3\sigma$) bound.}
\label{fig:RVV2}
\end{figure}

On the left panel of Figure~\ref{fig:RVV2}, we show the predictions for $\mu\to e\gamma$ and the electron EDM, $d_e$. Although they present important bounds, many points are still allowed. Furthermore, the future prospects for both observables shall probe almost completely the whole parameter space.

The central panel of Figure~\ref{fig:RVV2} shows the constraints by $\epsilon_K$ and $S_{\psi K_s}$. We see a very wide range of possible values, especially for the former. The quark sector shall then strongly constrain the parameter space. Nevertheless, this also means that this model has the potential to solve the CPV tensions in the $K$ and $B$ sectors.

Finally, the right panel of Figure~\ref{fig:RVV2} shows that, once the constraints of $\epsilon_K$ and $S_{\psi K_s}$ are taken, then the model easily satisfies the $\Delta M_B/\Delta M_{B_s}$ bounds, even if we just take the experimental errors. Also, we get a prediction for $S_{\psi\phi}$, with values somewhat larger than what the SM predicts.

\section{Conclusions}

We have shown that, if hints for SUSY are found with 2 fb$^{-1}$ of luminosity, then the interplay with flavour plays an important role in the disentanglement of the parameter space of each model.

More profound studies of this interplay, with an extension to 5 fb$^{-1}$, can be found in~\cite{Calibbi:2011dn}.

\end{document}